\documentclass[twocolumn,showpacs]{revtex4}
\usepackage{graphicx}
\usepackage{dcolumn}
\usepackage{bm}
\usepackage{amsmath}
\usepackage{amsfonts}
\usepackage{amssymb}
\providecommand{\U}[1]{\protect\rule{.1in}{.1in}}

\begin{document}
\title{Tailoring discrete quantum walk dynamics via extended initial conditions:
Towards homogeneous probability distributions}
\author{Germ\'{a}n J. de Valc\'{a}rcel$^{1}$, Eugenio Rold\'{a}n$^{1}$ and Alejandro
Romanelli$^{2}$} \affiliation{$^{1}$Departament d'\`{O}ptica,
Universitat de Val\`{e}ncia, Dr. Moliner 50, 46100-Burjassot, Spain,
EU} \affiliation{$^{2}$Instituto de F\'{\i}sica, Facultad de
Ingenier\'{\i}a, Universidad de la Rep\'{u}blica, C.C. 30, C.P.
11000, Montevideo, Uruguay}
\begin{abstract}
We study the evolution of initially extended distributions in the
coined quantum walk on the line by analyzing the dispersion relation
of the process and its associated wave equations. This allows us, in
particular, to devise an initially extended condition leading to a
uniform probability distribution whose width increases linearly with
time, with increasing homogeneity.
\end{abstract}
\maketitle
\textit{Introduction.--} The discrete, or coined, quantum walk (QW)
\cite{QW} is a process originally introduced as the quantum
counterpart of the classical random walk (RW). In both cases there
is a walker and a coin: at every time step the coin is tossed and
the walker moves depending on the toss output. In the RW the walker
moves to the right \textit{or} to the left, while in the QW, as the
walker and coin are quantum in nature, coherent superpositions
right/left and head/tail happen. This feature endows the QW with
outstanding properties, such as making the standard deviation of the
position of an initially localized walker grow linearly with time
$t$, unlike the RW in which this growth goes with $t^{1/2}$. This
has strong consequences in algorithmics and is one of the reasons
why QWs are receiving so much attention from the past decade.
However the relevance of QW's is being recognized to go beyond this
specific arena and, for example, some simple generalizations of the
standard QW have shown unsuspected connections with phenomena such
as Anderson localization \cite{Romanelli} and quantum chaos
\cite{Buerschaper,Wojcik}. Moreover, theoretical and experimental
studies evidence that the QW finds applications in outstanding
systems, such as Bose--Einstein condensates \cite{Chandrashekar06},
atoms in optical lattices \cite{Chandrashekar08,Science}, trapped
ions \cite{Schmitz,Zaehringer}, or optical devices
\cite{Knight,Bouwmester,Do,Waveguides,Fibre}, just to mention a few.
This enhances the relevance of the QW as it can constitute a means
for controlling the performance of those systems. Apart from the
discrete QW we consider here, continuous versions exist as well
\cite{Continuo}, whose relationship with the coined QW has been
discussed in \cite{Strauch06}.

Surprisingly enough even the simplest version of the discrete QW has
not been studied in all its extension. Specifically we refer to the
fact that in almost all studies up to date, the initial state of the
walker is assumed to be sharply localized at the line origin, with
few exceptions. In \cite{Strauch06} it was shown that for wider
initial distributions (an extended wavepacket), the evolution of the
wavepacket is Gaussian-like, not showing the two characteristic
outer peaks appearing in the probability distribution for more
sharply localized initial conditions. In \cite{Chandrashekar08} also
extended distributions, with top-hat profile, were considered in the
context of the superfluid-Mott insulator transition in optical
lattices, but no general conclusions were drawn on the influence of
these extended initial conditions on the long time state. It is this
issue that we address in this Letter, and the results we obtain open
the way to new types of distributions that the QW can exhibit, e.g.
virtually flat ones, with obvious impact in applications of this
process.

\textit{The coined QW on the line.--} In this QW the walker moves (at discrete
time steps $t\in%
\mathbb{N}
$) along a one-dimensional lattice of sites $x\in%
\mathbb{Z}
$, with a direction that depends on the state of the coin (with eigenstates
$R$ and $L$). The state of the total system at $\left(  x,t\right)  $ can be
expressed in the form,%
\begin{equation}
\left\vert \Psi_{x,t}\right\rangle =\operatorname{col}\left(  R_{x,t}%
,L_{x,t}\right)  ,
\end{equation}
where $R_{x,t}$ and $L_{x,t}$ are wave functions on the lattice. As
$\left\vert R_{x,t}\right\vert ^{2}$ and $\left\vert L_{x,t}\right\vert ^{2}$
have the meaning of probability of finding the walker at $\left(  x,t\right)
$ and the coin in state $R$ and $L$, respectively, the probability of finding
the walker at $\left(  x,t\right)  $ is
\begin{equation}
P_{x,t}=\left\langle \Psi_{x,t}\right.  \left\vert \Psi_{x,t}\right\rangle
=\left\vert R_{x,t}\right\vert ^{2}+\left\vert L_{x,t}\right\vert ^{2},
\label{prob}%
\end{equation}
and $\sum_{x}P_{x,t}=1$. The QW is ruled by a unitary map and a standard form
is \cite{Romanelli09}
\begin{subequations}
\label{map}%
\begin{align}
R_{x,t+1}  &  =R_{x+1,t}\cos\theta+L_{x+1,t}\sin\theta,\\
L_{x,t+1}  &  =R_{x-1,t}\sin\theta-L_{x-1,t}\cos\theta,
\end{align}
where $\theta\in\left[  0,\pi/2\right]  $ is a parameter defining the bias of
the coin toss ($\theta=\frac{\pi}{4}$ for an unbiased, or Hadamard, coin).

\textit{The dispersion relation and the group velocity.-- }Plane wave
solutions to (\ref{map}) exist in the form \cite{Ambainis} $\exp\left[
i\left(  kx-\omega^{\left(  s\right)  }t\right)  \right]  \left\vert \Phi
_{k}^{\left(  s\right)  }\right\rangle $, where $s=\pm$, $k\in\left[
-\pi,+\pi\right]  $, $\omega^{\left(  +\right)  }=\omega$, $\omega^{\left(
-\right)  }=\pi-\omega$,%
\end{subequations}
\begin{align}
\omega &  =-\arcsin\left(  \cos\theta\sin k\right)  \in\left[  -\tfrac{\pi}%
{2},\tfrac{\pi}{2}\right]  ,\label{disp}\\
\left\vert \Phi_{k}^{\left(  \pm\right)  }\right\rangle  &  =\mathcal{N}_{\pm
}\operatorname{col}\left(  \cos\theta\cos k\pm\cos\omega,e^{-ik}\sin
\theta\right)  ,\label{eigenvectors}%
\end{align}
and $\mathcal{N}_{\pm}$ is a normalization factor (any $\mathcal{N}$ will have
this meaning in the following). The dispersion relation (\ref{disp}) is
represented in Fig. 1 together with the group velocity $v_{\mathrm{g}%
}^{\left(  +\right)  }\left(  k\right)  =d\omega/dk$ associated with
$\left\vert \Phi_{k}^{\left(  +\right)  }\right\rangle $ \cite{vg}.%
\begin{figure}[th]
\begin{center}
\includegraphics[scale=0.35]{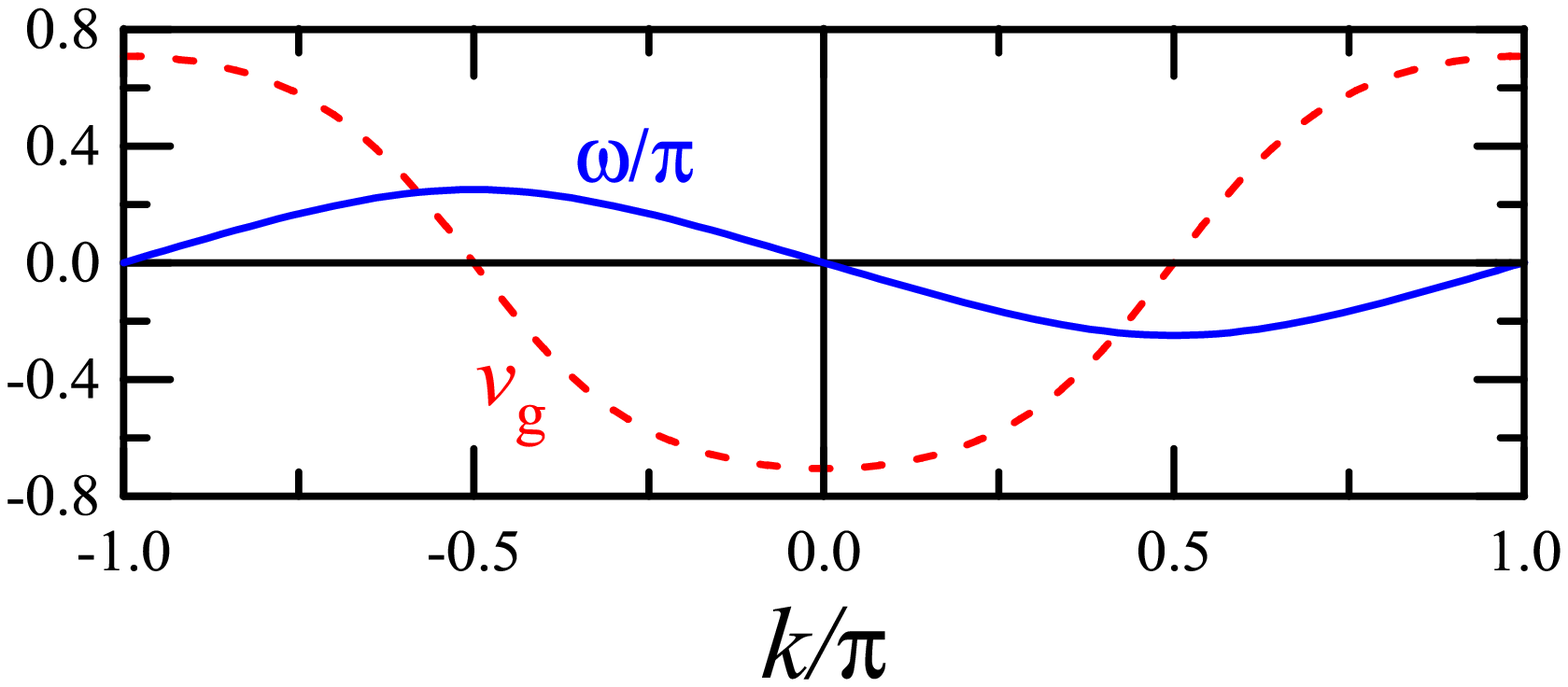}
\end{center}
\caption{(Colour online) Dispersion relation (full line) as given by
Eq. (\ref{disp}), and corresponding group velocity (dashed line).
$\theta=\pi/4$.} \label{f2}
\end{figure}
The QW group velocity has been used for determining hitting times
\cite{Kempf}, and it will allow us to make simple but relevant predictions
about the QW dynamics when the initial state is a wavepacket close to some of
the eigensolutions above, say $\left\vert \Psi_{x,0}\right\rangle
=f_{x}^{\left(  s\right)  }e^{ik_{0}x}\left\vert \Phi_{k_{0}}^{\left(
s\right)  }\right\rangle $ with $f_{x}^{\left(  s\right)  }$ a smooth
envelope. In that case, as in any linear wave system, one must expect that, to
the leading order, the group velocity govern its propagation. Hence if
$k_{0}=\pm\pi/2$ a sufficiently extended wavepacket should stay at rest
because $v_{\mathrm{g}}^{\left(  s\right)  }\left(  \pm\pi/2\right)  =0$,
while if $k_{0}=0$ it should move with maximum velocity $v_{\mathrm{g}%
}^{\left(  s\right)  }\left(  0\right)  =-s\cos\theta$. If the
initial condition projects onto both $\left\vert
\Phi_{k_{0}=0}^{\left(  \pm\right) }\right\rangle $ we must expect
that the initial wavepacket splits into two, moving at opposite
velocities given by $\mp\cos\theta$. Numerical simulations of the QW
(with Gaussian initial conditions, see below) confirm these notable
effects and tell us that the dispersion relation is a powerful tool
for predicting QW dynamics \cite{Kempf}. As we demonstrate in the
next sections the dispersion relation (\ref{disp}) controls not only
the velocity of the wavepacket, but also the evolution of its shape
as time runs, what will allow us making interesting predictions.

\textit{Broad initial distributions: Wave equations in the continuum limit.--}
The goal of this section is to find a wave equation for the wavepacket
envelope with the help of discrete Fourier analysis. Given a function $f_{x}$
on integers $x\in%
\mathbb{Z}
$, one can define its discrete Fourier transform (DFT) as $\tilde{f}_{k}%
=\sum_{x}f_{x}e^{-ikx}$, which can be inverted as $f_{x}=\left(  2\pi\right)
^{-1}\int_{-\pi}^{+\pi}dk\tilde{f}_{k}e^{ikx}$. Applying this DFT to the map
(\ref{map}) it is straightforward to get an explicit solution to the QW given
an arbitrary initial condition $\left\vert \Psi_{x,0}\right\rangle $. The
result is $\left\vert \Psi_{x,t}\right\rangle =\sum_{s=\pm1}\left\vert
\Psi_{x,t}^{\left(  s\right)  }\right\rangle $, where%
\begin{equation}
\left\vert \Psi_{x,t}^{\left(  s\right)  }\right\rangle =\int_{-\pi}^{+\pi
}\frac{dk}{2\pi}e^{i\left(  kx-\omega^{\left(  s\right)  }t\right)
}\left\vert \Phi_{k}^{\left(  s\right)  }\right\rangle \left\langle \Phi
_{k}^{\left(  s\right)  }\right.  \left\vert \tilde{\Psi}_{k,0}\right\rangle ,
\label{exact}%
\end{equation}
and $\left\vert \tilde{\Psi}_{k,0}\right\rangle =\sum_{x}e^{-ikx}\left\vert
\Psi_{x,0}\right\rangle $.

As stated, we are interested in initial conditions of the form $\left\vert
\Psi_{x,0}\right\rangle =\sum_{s=\pm1}f_{x}^{\left(  s\right)  }e^{ik_{0}%
x}\left\vert \Phi_{k_{0}}^{\left(  s\right)  }\right\rangle $, where the
envelopes $f_{x}^{\left(  s\right)  }$ vary smoothly on $x$, and $k_{0}$ is a
(carrier) wave number. Then $\left\vert \tilde{\Psi}_{k,0}\right\rangle
=\sum_{s=\pm1}\tilde{f}_{k-k_{0}}^{\left(  s\right)  }\left\vert \Phi_{k_{0}%
}^{\left(  s\right)  }\right\rangle $, which is peaked around $k=k_{0}$ as
$\tilde{f}_{k-k_{0}}^{\left(  s\right)  }$ is peaked around $k-k_{0}=0$ (low
frequency envelope). In such cases Eq. (\ref{exact}) can be written as%
\begin{align}
\left\vert \Psi_{x,t}^{\left(  s\right)  }\right\rangle  &  =e^{i\left(
k_{0}x-\omega_{0}^{\left(  s\right)  }t\right)  }F_{s}\left(  x,t\right)
\left\vert \Phi_{k_{0}}^{\left(  s\right)  }\right\rangle +\mathcal{O}\left(
\Delta k\right)  ,\\
F_{s}\left(  x,t\right)   &  =\int_{-\pi}^{+\pi}\frac{dK}{2\pi}\tilde{f}%
_{K}^{\left(  s\right)  }e^{i\left(  Kx-s\Omega t\right)  },\label{Fs}%
\end{align}
where $\Delta k$ is the width of $\tilde{f}_{k}^{\left(  s\right)  }$,
$K=k-k_{0}$, $\Omega=\omega-\omega_{0}$, and we did not modify the limits of
the integral because of the assumed smallness of $\Delta k$. We have
introduced two wave functions, $F_{\pm}\left(  x,t\right)  $, in terms of
which $P_{x,t}=\sum_{s=\pm1}\left\vert F_{s}\left(  x,t\right)  \right\vert
^{2}+\mathcal{O}\left(  \Delta k\right)  $. We let $F_{s}\left(  x,t\right)  $
be defined on the reals, as there is nothing against that in Eq. (\ref{Fs}),
so that it is straightforward setting a wave equation from that equation,%
\begin{equation}
i\partial_{t}F_{s}\left(  x,t\right)  =-is\omega_{1}\partial_{x}F_{s}%
-\tfrac{1}{2}s\omega_{2}\partial_{x}^{2}F_{s}+\cdots,\label{wave2}%
\end{equation}
after Taylor expanding $\Omega$ around $k_{0}$, and where $\omega_{n}=\left(
d^{n}\omega/dk^{n}\right)  _{k=k0}$. This equation is to be solved under the
initial condition $F_{s}\left(  x,0\right)  =f_{x}^{\left(  s\right)  }$ at
integer $x$ \cite{cin}.

Eq. (\ref{wave2}) is a main result of this Letter. It evidences the role
played by the dispersion relation (\ref{disp}) as anticipated: For
distributions whose DFT is centered around some $k_{0}$, the local variations
of $\omega$ around $k_{0}$ determine the type of wave equation controlling the
QW dynamics. The first term on the rhs gives the group velocity, already
discussed, the second accounts for diffraction, and so on.

\textit{Application to Gaussian initial distributions.--} Two cases of
interest of Eq. (\ref{wave2}) are analyzed next, corresponding to $k_{0}%
=0,\pi/2$ as suggested by the analysis of the dispersion relation. First, the
case $k_{0}=0$ yields, to the leading order,%
\begin{equation}
\partial_{t}F_{s}=\left(  s\cos\theta\right)  \partial_{x}F_{s}. \label{lin}%
\end{equation}
According to (\ref{lin}), if the (broad) initial condition projects onto both
eigenspinors $\left\vert \Phi_{0}^{\left(  \pm\right)  }\right\rangle $, two
wavepackets (whose height will depend on the projections $\left\langle
\Phi_{0}^{\left(  \pm\right)  }\right.  \left\vert \Psi_{x,0}\right\rangle $)
will propagate without distortion at equal but opposite velocities given by
$v_{\mathrm{g}}^{\left(  s\right)  }\left(  0\right)  =-s\cos\theta$ as
commented above. We have checked this prediction in the original QW map
(\ref{map}) with initial states of the form $\left\vert \Psi_{x,0}%
\right\rangle =\mathcal{N}\exp\left[  -\frac{1}{2}\left(  x/\sigma_{0}\right)
^{2}\right]  \left\vert C\right\rangle $, which is a Gaussian of width
$\sigma_{0}$. The state $\left\vert C\right\rangle $ of the coin (taken equal
at any site) controls the projections $\left\langle \Phi_{0}^{\left(
\pm\right)  }\right.  \left\vert \Psi_{x,0}\right\rangle $. The distortionless
propagation at a velocity $v_{\mathrm{g}}^{\left(  s\right)  }\left(
0\right)  $ is observed in excellent agreement with the prediction even for
moderate values of $\sigma_{0}$. Nevertheless, in all cases, a deformation of
the wavepackets is visible after some running time, the longer the wider the
initial distribution is. This deformation is controlled by the next, third
order derivative term, in which case one has an equation similar to that
derived in \cite{Knight}, where the role of the third order derivative was
analyzed. This type of approximation was shown to be quite good even for
localized initial conditions, where the truncation of the dispersion relation
is not so well justified as the initial condition projects over all $k$ values.

The case $k_{0}=\pi/2$ is more interesting for our purposes. It is described,
to the leading order, by%
\begin{equation}
i\partial_{t}F_{s}=-\frac{s}{2\tan\theta}\partial_{x}^{2}F_{s}%
,\label{paraxial}%
\end{equation}
which is analogous to the Schr\"{o}dinger equation as well as to the equation
of paraxial optical diffraction, and the pulse propagation equation in linear
optical fibers. The solution to (\ref{paraxial}) under Gaussian initial
condition is $F_{s}\left(  x,t\right)  =\mathcal{N}\exp\left[  -\frac{1}%
{2}\left(  x-x_{0}\right)  ^{2}/\left(  \sigma_{0}q_{s}\right)  ^{2}\right]
$, where $\sigma_{0}$ is the initial width and $x_{0}$ is center of the
distribution (that remains constant as the group velocity is null in the
present case), and the complex parameter $q_{s}\left(  t\right)
=\sqrt{1+ist/\sigma_{0}^{2}\tan\theta}$. The probability $P_{s}\left(
x,t\right)  =\mathcal{N}\exp\left[  -\left(  x-x_{0}\right)  ^{2}/\left(
\sigma_{0}w\right)  ^{2}\right]  $ thus remains Gaussian with $w\left(
t\right)  =\sqrt{1+\left(  t/\sigma_{0}^{2}\tan\theta\right)  ^{2}}$ the width
of the distribution relative to its initial width $\sigma_{0}$, which grows
linearly with time as soon as $t\gtrsim5\sigma_{0}^{2}\tan\theta$ (note that
these results are independent of $s$). Again, numerical simulations of
(\ref{map}), now with $\left\vert \Psi_{x,0}\right\rangle =\mathcal{N}%
\exp\left[  -\left(  x/\sigma_{0}\right)  ^{2}/2+i\pi x/2\right]
\left\vert C\right\rangle $ are in excellent agreement with the
analytical predictions.
\begin{figure}[th]
\begin{center}
\includegraphics[scale=0.35, angle=-90]{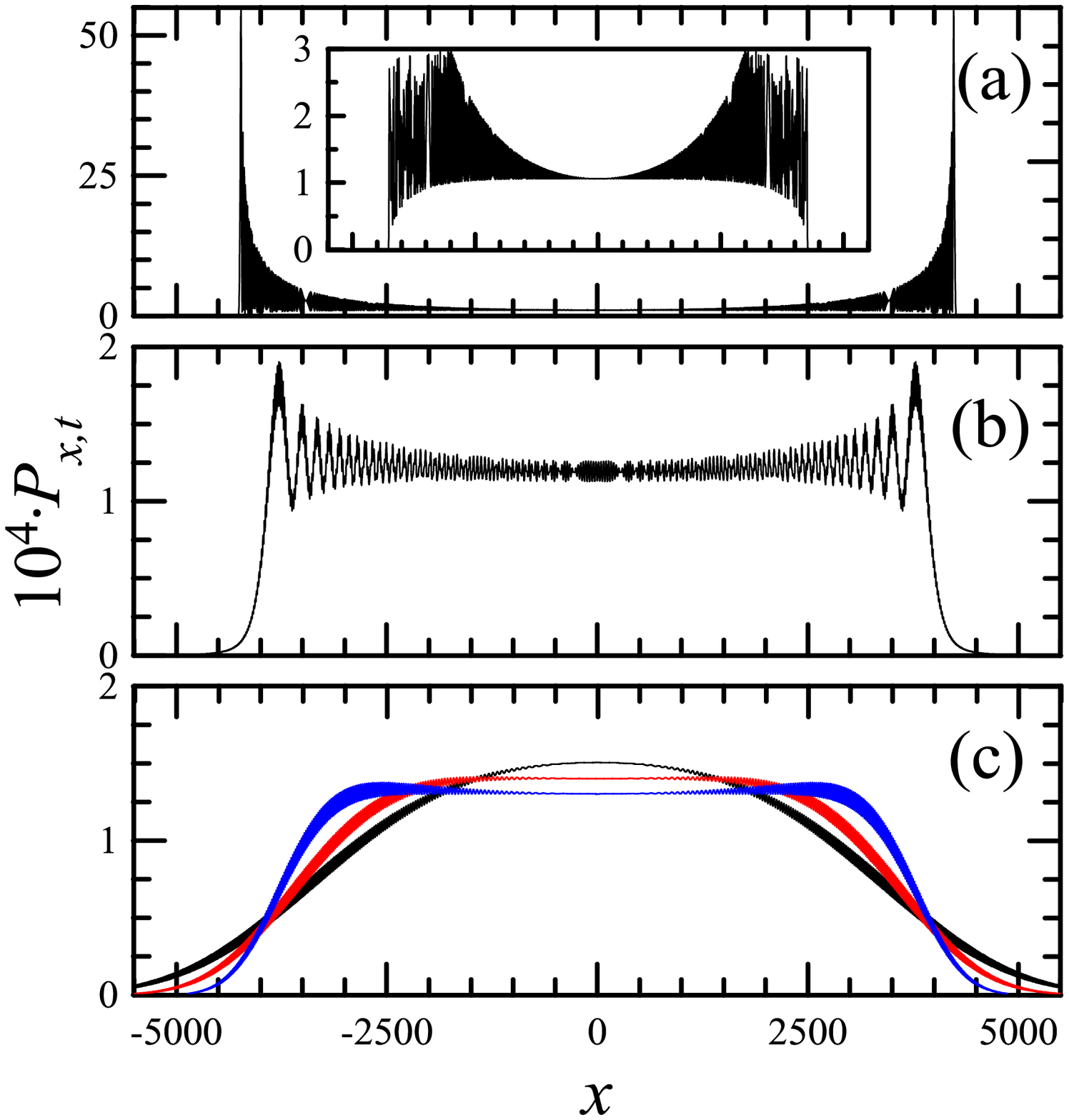}
\end{center}
\caption{Probability distributions for $\theta=\pi/4$ and different
initial
conditions. In (a) the initial condition is $\left\vert \Psi_{x,0}%
\right\rangle =\delta_{x,0}\left\vert \Phi_{\pi/2}^{\left(  +\right)
}\right\rangle $, fully localized at the origin, and the time run is
$t=6\times10^{3}$; the inset shows a magnification (odd sites have
zero occupation probability). In (b) $\left\vert
\Psi_{x,0}\right\rangle =\mathcal{N}\exp\left(  i\pi x/2\right)
\operatorname{sinc}\left( x/\sigma_{0}\right)  \left\vert
\Phi_{\pi/2}^{\left(  +\right)  }\right\rangle
$ is used while in (c) additional Gaussians (of widths $\sigma_{\mathrm{G}%
}=1.1\sigma_{0},2\sigma_{0}$, and $3\sigma_{0}$, from top to bottom)
multiply
the initial condition; $\sigma_{0}=15$ and the time run is $t=20\times10^{3}%
$.} \label{f2}
\end{figure}

\textit{Achieving homogeneous distributions.--} In QW's a most
desired result is that the probability distribution be as uniform as
possible after a time. A hint towards that goal is given by light
paraxial diffraction theory --that the QW follows for $k_{0}=\pi/2$,
Eq. (\ref{paraxial}): It is a textbook result that the far field
corresponding to a light amplitude distribution
$\operatorname{sinc}\left(  x\right)  =\sin\left(  \pi x\right)
/\left(  \pi x\right)  $ is remarkably homogeneous within a certain
spatial region \cite{Goodman}.

In order to gain insight into the problem we look at the solution to Eq.
(\ref{paraxial}) under an initial condition $F\left(  x,0\right)
=\mathcal{N}\operatorname{sinc}\left(  x/\sigma_{0}\right)  $. This solution
is obtainable by Fourier transformation of the spatial coordinate
\cite{Agrawal}, $F_{s}\left(  x,t\right)  =\int_{-\infty}^{+\infty}dk\tilde
{F}_{s}\left(  k,0\right)  \exp\left(  ikx-isk^{2}t/2\tan\theta\right)  $,
with $\tilde{F}_{s}\left(  k,0\right)  =\left(  2\pi\right)  ^{-1}%
\int_{-\infty}^{+\infty}dxF_{s}\left(  x,0\right)  \exp\left(  -ikx\right)  $.
In our case $\tilde{F}_{s}\left(  k,0\right)  =\left(  2\pi\right)  ^{-1}%
\sqrt{\sigma_{0}}\operatorname{rect}\left(  \sigma_{0}k/\pi\right)  $, where
$\operatorname{rect}\left(  a\right)  =1$ if $\left\vert a\right\vert <1$ and
$0$ otherwise. Hence $F_{s}\left(  x,t\right)  =\left(  2\pi\right)
^{-1}\sqrt{\sigma_{0}}\int_{-\pi/\sigma_{0}}^{+\pi/\sigma_{0}}dk\exp\left[
it\phi_{s}\left(  k,\xi\right)  \right]  $, with $\phi_{s}\left(
k,\xi\right)  =k\xi-sk^{2}/2\tan\theta$ and $\xi=x/t$.

As we are interested in the behavior of $F_{s}$ at long times we can use the
method of stationary phase \cite{stat phase} to evaluate the leading
dependence of the integral on $\left(  x,t\right)  $. It is straightforward to
get $F_{s}\left(  x,t\right)  =w\left(  t\right)  ^{-1/2}\exp\left[  -i\left(
s\pi/4+x^{2}\tan\theta/2t\right)  \right]  $ with\ $w\left(  t\right)
=2\pi\left(  \sigma_{0}\tan\theta\right)  ^{-1}t$. Hence the asymptotic
analysis predicts a uniform probability $P_{s}\left(  x,t\right)  =w\left(
t\right)  \operatorname{rect}\left[  2x/w\left(  t\right)  \right]  $, inside
a segment of width $w\left(  t\right)  $. Thus we should expect, after a
transient time (that can be estimated as $t\gtrsim2\sigma_{0}^{2}\tan\theta$),
a flat distribution whose width increases \textit{linearly} with time (its
standard deviation is $w\left(  t\right)  /\sqrt{12}$).

Inspired by this result we consider the actual QW initial condition
$\left\vert \Psi_{x,0}\right\rangle =\mathcal{N}\exp\left(  i\pi x/2\right)
\operatorname{sinc}\left(  x/\sigma_{0}\right)  \left\vert \Phi_{\pi
/2}^{\left(  +\right)  }\right\rangle $ with $\sigma_{0}$ the initial width.
Figure 2(b) shows results of the simulation of (\ref{map}), which are in
qualitative agreement with the discussion above: A quite uniform distribution
is attained. For comparison, the well known result corresponding to an
initially localized distribution is shown in Fig. 2(a). Main differences are
the improved degree of uniformity in (b), which is free from the large outer
peaks in (a), and the fact that in (a) even/odd sites have null occupation
probability at odd/even times unlike in (b). There is however a disgusting
feature in (b), namely the high frequency ripples that appear at the plateau.
Nevertheless the situation can be improved by multiplying the initial
condition by a Gaussian of convenient width (this is a kind of smoothing
processing, typical of optical diffraction \cite{Goodman}), as shown in Figure
2(c), which is another main result of this Letter: Almost uniform
distributions (even reaching a top hat profile) can be obtained in the QW by
making a judicious choice of the initial condition. We want to stress that
these homogeneous distributions are so after a short transient, their
homogeneity increasing with time.

The achievement of QW homogenous distributions is a most desired property.
From the very beginning the relatively high homogeneity of the probability
distribution of the QW corresponding to a localized initial condition has been
considered as a positive quality of this process for information purposes. In
\cite{Kendon} the presence of some decoherence in the process was considered
to be beneficial because it leads to more homogeneous distributions \textit{at
a special time}. In this sense our finding may have relevance as we have seen
that a judicious initial condition helps in achieving distributions with much
larger and permanent homogeneity than the initially localized case, even
including decoherence, as Fig. 2(c) clearly demonstrates.

\textit{Conclusions and discussion.--} In this work we have studied the
influence of the initial condition on the discrete QW on the line guided by
the QW dispersion relation and by its associated wave equations. Specifically
we have considered the evolution of initially extended probability
distributions. We have shown that sufficiently wide Gaussian initial
distributions propagate without distortion and small width increase at
velocities that can be tuned, including null velocity, with a proper choice of
the phases along the initial distribution as given by the phase factor
$\exp\left(  ik_{0}x\right)  $. We have devised as well an initial condition
that leads, after a transient, to a homogeneous distribution whose width
increases with time remaining highly homogeneous at any later time. This
result, Fig. 2(c), is a main result of this paper.

We further mention that any behavior of light diffraction (or linear pulse
propagation) can be transferred to the QW in the case of the resting
probability distributions ($k_{0}=\pi/2$), as its continuous limit, Eq.
(\ref{paraxial}), is nothing but the paraxial diffraction equation. For
instance, a light pattern replicates at specific planes when it is periodic,
the so called Talbot effect \cite{Goodman}. In our case, if $F_{s}\left(
x+\lambda,0\right)  =F_{s}\left(  x,0\right)  $, where $\lambda$ is the
spatial period, then $\left\vert F_{s}\left(  x,nT\right)  \right\vert
^{2}=\left\vert F_{s}\left(  x,0\right)  \right\vert ^{2}$, where $T=\left(
2\pi\right)  ^{-1}\lambda^{2}\tan\theta$ is the Talbot period and $n$ is
integer. Other optical effects, such as those of lenses, can also be mimicked
by introducing quadratic phase factors \cite{Goodman} in the initial
condition. It is just the choice of that condition that can make any paraxial,
linear optical phenomenon be reproduced with the QW, what can be useful for
special implementations of this rich process.

This work has been supported by the Spanish Government and the European Union
FEDER through Project FIS2008-06024-C03-01. A.R. acknowledges financial
support from PEDECIBA, ANII, Universitat de Val\`{e}ncia and Generalitat Valenciana.


\begin{thebibliography}{99}
\bibitem {vg}The group velocity corresponding to $\left\vert \boldsymbol{\Phi
}_{k}^{\left(  -\right)  }\right\rangle $ is $v_{\mathrm{g}}^{\left(
-\right)  }\left(  k\right)  =d\omega^{\left(  -\right)  }/dk=-d\omega/dk$.
\bibitem {cin}According to (\ref{Fs}) $F_{s}\left(  x,0\right)  =\left(
2\pi\right)  ^{-1}\int_{-\pi}^{+\pi}dK\tilde{f}_{K}^{\left(  s\right)  }%
\exp\left(  iKx\right)  =\sum_{y}f_{y}^{\left(  s\right)  }\operatorname{sinc}%
\left(  x-y\right)  $, where $\operatorname{sinc}\left(  x\right)
=\sin\left(  \pi x\right)  /\left(  \pi x\right)  $, and at integer $x$
$\operatorname{sinc}\left(  x-y\right)  =\delta_{x,y}$. In general the
relation between $f_{x}^{\left(  s\right)  }$ and $F_{s}\left(  x,0\right)  $
is not invertible, what means that not all the choices for the latter
correspond to a true $f_{x}^{\left(  s\right)  }$. However, any choice of
$F_{s}\left(  x,0\right)  $ varying smoothly on the unit spatial scale
corresponds to a true $f_{x}^{\left(  s\right)  }=F_{s}\left(  x,0\right)  $
at integer $x$.
\end{thebibliography}
\end{document}